\documentclass[dvips,12pt,a4paper]{article}
\usepackage{a4wide}
\usepackage{epsfig}
\usepackage{amsmath}
\usepackage{latexsym}

 \newlength{\absize}
\allowdisplaybreaks 

\catcode`@=11
\def\citer{\@ifnextchar [{\@tempswatrue\@citexr}{\@tempswafalse\@citexr[]}}
 
\def\@citexr[#1]#2{\if@filesw\immediate\write\@auxout{\string\citation{#2}}\fi
  \def\@citea{}\@cite{\@for\@citeb:=#2\do
    {\@citea\def\@citea{--\penalty\@m}\@ifundefined
       {b@\@citeb}{{\bf ?}\@warning
       {Citation `\@citeb' on page \thepage \space undefined}}%
\hbox{\csname b@\@citeb\endcsname}}}{#1}}
\catcode`@=12
\begin{document}
  \thispagestyle{empty}
  \pagestyle{empty}
  \renewcommand{\thefootnote}{\fnsymbol{footnote}}
\newpage\normalsize
    \pagestyle{plain}
    \setlength{\baselineskip}{4ex}\par
    \setcounter{footnote}{0}
    \renewcommand{\thefootnote}{\arabic{footnote}}
\newcommand{\preprint}[1]{%
  \begin{flushright}
    \setlength{\baselineskip}{3ex} #1
  \end{flushright}}
\renewcommand{\title}[1]{%
  \begin{center}
    \LARGE #1
  \end{center}\par}
\renewcommand{\author}[1]{%
  \vspace{2ex}
  {\Large
   \begin{center}
     \setlength{\baselineskip}{3ex} #1 \par
   \end{center}}}
\renewcommand{\thanks}[1]{\footnote{#1}}
\renewcommand{\abstract}[1]{%
  \vspace{2ex}
  \normalsize
  \begin{center}
    \centerline{\bf Abstract}\par
    \vspace{2ex}
    \parbox{\absize}{#1\setlength{\baselineskip}{2.5ex}\par}
  \end{center}}

\begin{flushright}
{\setlength{\baselineskip}{2ex}\par
} 
\end{flushright}
\vspace*{1truecm}
\vspace*{4mm}
\vfill
\title{The $\sigma $ and $\rho $ in $D$ and $B$ decays}
\vfill
\author{
Nello Paver$^{a,b}$ {\rm and} 
Riazuddin$^{c,d}$}

\begin{center}
$^a$ Dipartimento di Fisica Teorica, Universit\`a di Trieste  \\
$^b$ Istituto Nazionale di Fisica Nucleare, Sezione di Trieste, 
Trieste, Italy \\
$^c$ The Abdus Salam International Centre for Theoretical Physics, Trieste, 
Italy  \\ 
$^d$ National Center for Physics, 
Quaid-i-Azam University, Islamabad 45320, Pakistan
\end{center}
\vspace*{2truecm}
\vfill
\abstract
{We study the $D^{+}\rightarrow \sigma \pi ^{+},$ $D^{+}\rightarrow \rho
^0\pi ^{+},$ $B^{-}\rightarrow \sigma \pi ^{-},$ $B^{-}\rightarrow \rho
^0\pi ^{-}$ and $\bar{B}^0\rightarrow \rho ^{\pm }\pi ^{\mp }$ 
decays in a 
valence quark triangle model, incorporating chiral symmetries. We find a 
good agreement with recent experimental data for $D^{+}\rightarrow \sigma
\pi ^{+}$ and for $D^{+}\rightarrow \rho ^0\pi ^{+}$. We point out that a 
long-distance contribution due to the axial vector $a_1$ meson pole, 
calculated by using chiral symmetry, can be relevant to explain 
$D^{+}\rightarrow \rho^0\pi ^{+}$ and for lowering the ratio 
\[{\cal R}=\frac{{\cal B}\left( \bar{B}^0\rightarrow \rho ^{\pm }\pi ^{\mp
}\right) }{{\cal B}\left( B^{-}\rightarrow \rho ^0\pi ^{-}\right) } 
\]
to be consistent with its phenomenological determination, within the large
experimental uncertainity.}

\vspace*{20mm}
\setcounter{footnote}{0}
\vfill

\newpage
    \setcounter{footnote}{0}
    \renewcommand{\thefootnote}{\arabic{footnote}}
    \setcounter{page}{1}

\section{Introduction}


Recently there has been a revival of interest 
\citer{babukhadia,tornqvist} in a broad 
scalar-isoscalar light $\pi \pi $ resonance, the $\sigma $ meson, which 
has been controversial for a long time. It has appeared in the Reviews of 
Particle Physics \cite{groom}, as a broad resonance under the entry 
$f_0\left(400-1200\right) $ or $\sigma $. The E791 collaboration measurement 
of the $D^{+}\rightarrow 3\pi $ rate provides an evidence for 
a scalar resonance $\sigma$ having mass $m_\sigma =478\pm 24$ MeV and width 
$\Gamma _\sigma
=324\pm 41$ MeV; the $\sigma $ is seen as a dominant peak leading to a fit
in which 46\% of the rate occurs via $D^{+}\rightarrow \sigma \pi^+ $ while
33\% of the rate occurs via 
$D^{+}\rightarrow \rho ^0\pi ^{+}$ \cite{aitala}. There has been 
considerable interest in explaining these rates \cite{gatto}. 

The effective weak Hamiltonian for the above decays can be written as 
\cite{bauer} 
\begin{equation}
H_{\text{eff}}=\frac{G_F}{\sqrt{2}}V_{cd}^{*}V_{ud}\left\{ a_1\left( \bar{d}
c\right) _{\text{V}-\text{A}}\left( \bar{u}d\right) _{\text{V}-\text{A}
}+a_2\left( \bar{d}d\right) _{\text{V}-\text{A}}\left( \bar{u}c\right) _{
\text{V}-\text{A}}\right\} ,  \label{1}
\end{equation}
where, in the factorization ansatz, $a_1=1.10\pm 0.05$ and $a_2=-0.49\pm
0.04$ fitted with $D$-decays \cite{bauer,neubert}. In this ansatz the 
relevant transition matrix elements are given as $\left[ A_\mu =\bar{d}
\gamma _\mu \gamma _5c\text{, }V_\mu =\bar{u}\gamma _\mu c\right] $: 
\begin{eqnarray}
\left\langle \sigma \left( k\right) \pi ^{+}\left( q\right) \left| H_{\text{
eff}}\right| D^{+}\left( p\right) \right\rangle  &=&\frac{G_F}{\sqrt{2}}
V_{cd}^{*}V_{ud}a_1f_\pi \left( -iq^\mu \right) \left\langle \sigma \left(
k\right) \left| A_\mu \right| D^{+}\left( p\right) \right\rangle,\label{2} \\
\left\langle \rho \left( k\right) \pi ^{+}\left( q\right) \left| H_{\text{eff
}}\right| D^{+}\left( p\right) \right\rangle  &=&\frac{G_F}{\sqrt{2}}
V_{cd}^{*}V_{ud}\left[ a_1f_\pi \left( -iq^\mu \right) \left\langle \rho
\left( k\right) \left| A_\mu \right| D^{+}\left( p\right) \right\rangle
\right.   \nonumber \\
&&\left. +a_2\left( -\frac{f_\rho }{\sqrt{2}}\right) \epsilon ^{*\mu
}\left\langle \pi \left( q\right) \left| V_\mu \right| D^{+}\left( p\right)
\right\rangle \right]\hskip 2pt .\label{3} 
\end{eqnarray}

The problem thus reduces to evaluating the form factors 
\begin{eqnarray}
A_\mu ^{D\sigma } &=&\left\langle \sigma \left( k\right) \left| A_\mu
\right| D^{+}\left( p\right) \right\rangle  \nonumber \\
&=&i\left[ G_{+}\left( q^2\right) \left( p+k\right) _\mu +G_{-}\left(
q^2\right) \left( p-k\right) _\mu \right]  \nonumber \\
&=&i\left[ \left( \frac{m_D^2-m_\sigma ^2}{q^2}\right) q_\mu G_0\left(
q^2\right) +\left( \left( p+k\right) _\mu -\frac{m_D^2-m_\sigma ^2}{q^2}
q_\mu \right) G_1\left( q^2\right) \right] , \label{4}  \\
A_\mu ^{D\rho } &=&\left\langle \rho ^0\left( k\right) \left| A_\mu \right|
D^{+}\left( p\right) \right\rangle  \nonumber \\
&=&i\left [ \left( \epsilon _\mu ^{*} -q_\mu\frac{\epsilon^{*}\cdot q}
{q^2}\right) \left(
m_\rho +m_D\right) A_1\left( q^2\right) -\left( \left( p+k\right) _\mu -
\frac{m_D^2-m_\rho ^2}{q^2}q_\mu \right) \epsilon ^{*}\cdot q\frac{
A_2\left( q^2\right) }{m_\rho +m_D}\right.  \nonumber \\
&&\left. +q_\mu \epsilon ^{*}\cdot q\frac{2m_\rho }{q^2}A_0\left( q^2\right)
\right ] , \label{5}   \\
V_\mu ^{D\pi } &=&\left\langle \pi ^{+}\left( q\right) \left| V_\mu \right|
D^{+}\left( p\right) \right\rangle \nonumber \\
&=&F_{+}\left( k^2\right) \left( p+q\right) _\mu +F_{-}\left( k^2\right)
\left( p-q\right) _\mu\hskip 2pt .\label{6}
\end{eqnarray}
Thus we obtain: 
\begin{eqnarray}
\left\langle \sigma \left( k\right) \pi ^{+}\left( q\right) \left| H_{\text{
eff}}\right| D^{+}\left( p\right) \right\rangle &=&\frac{G_F}{\sqrt{2}}
V_{cd}^{*}V_{ud}a_1f_\pi \left( m_D^2-m_\sigma ^2\right) 
\left[ G_0^{D\sigma}\left( m_\pi ^2\right) \right], \label{7}   \\
\left\langle \rho \left( k\right) \pi ^{+}\left( q\right) \left| H_{\text{eff
}}\right| D^{+}\left( p\right) \right\rangle &=&\frac{G_F}{\sqrt{2}}
V_{cd}^{*}V_{ud}a_1f_\pi \left( 2m_\rho \right) \left( \epsilon ^{*}\cdot
q\right) \biggl[ A_0^{D\rho }\left( m_\pi ^2\right) \biggr.  \nonumber \\
&&\left. 
-\frac{a_2}{a_1}\hskip 2pt\frac{f_\rho}{\sqrt{2}f_\pi m_\rho}\hskip 2pt
F_{+}^{D\pi}\left( m_\rho ^2\right) \right]\hskip 2pt .\label{8}
\end{eqnarray}

We evaluate the above form factors $G_0^{D\sigma }$ and $A_0^{D\rho }$ in 
the model based on the constituent quark ``triangle'' graph of Fig. 1. It is 
in this respect that we differ from the calculation in Ref. \cite{gatto}. 
Moreover, we take into account the ``long-distance'' contribution coming 
through the $a_1^{+}$-pole shown in Fig. 2, which has not been previously 
considered. Here, the weak vertex in the factorization ansatz can be 
expressed as
\begin{equation} 
\left\langle a_1\left| H_w\right| D^{+}\right\rangle =\frac{G_F}{\sqrt{2}}
V_{cd}^{*}V_{ud}a_1f_{a_1}\left( if_Dp^\mu \right) ,  \label{9}
\end{equation}
with $f_D$ the leptonic constant of the $D$ meson, while the strong vertices 
are defined by 
\begin{equation}  
\left\langle \left. \sigma \left( k\right) \pi ^{+}\left( q\right) \right|
a_1\left( p\right) \right\rangle =i\frac 12\gamma _{a_1\sigma \pi }\eta
\cdot \left( q-k\right) ,  \label{10}
\end{equation}
corresponding to the Lagrangian $\left( \sigma \partial _\mu \pi 
-\pi\partial _\mu \sigma \right) \cdot a_1^\mu $. Moreover, 
\begin{equation}
\left\langle \left. \rho \left( k\right) \pi ^{+}\left( q\right) 
\right|a_1\left( p\right) \right\rangle =i\left( m_{a_1}^2-m_\rho 
^2\right) \eta\cdot \epsilon ^{*}f_{a_1\rho \pi },  \label{11}
\end{equation}
where we have neglected the $D$-wave coupling $g_{a_1\rho \pi }$, 
for which there is evidence to be negligible \cite{groom}: 
\begin{equation} 
\frac{D\text{-wave amplitude}}{S\text{-wave amplitude}}
=-0.107\pm 0.016.\label{12}
\end{equation}

The above considerations can be easily extended to $B^{-}\rightarrow \rho
^0\pi ^{-}$ and $\bar{B}^0\rightarrow \rho ^{\pm }\pi ^{\mp }$, where the $
a_1^{-} $ contributes to $B^{-}\rightarrow \rho ^0\pi ^{-}$ and only in a 
negligible way (being proportional to $a_2$) to $\bar{B}^0\rightarrow \rho
^{\pm }\pi ^{\mp }$. In principle, this provides a mechanism to lower the 
value of the ratio 
\[
{\cal R}=\frac{{\cal B}\left( \bar{B}^0\rightarrow \rho ^{\pm }\pi ^{\mp
}\right) }{{\cal B}\left( B^{-}\rightarrow \rho ^0\pi ^{-}\right) }. 
\]
Previous theoretical estimates computed in the simple factorization ansatz 
of Ref.~\cite{bauer} tend to give this ratio much larger than its  
experimental value: $\left(2.65\pm 1.8\right) $ or $\left( 2.0\pm 1.3\right) 
$ determined, respectively, from the measured indicated branching ratios 
in \cite{jessop} and \cite{aubert}. Recent efforts to understand the size 
of this ratio have been published, {\it e.g.}, 
in Refs.\cite{deandrea} and \citer{gao,deandrea2}.

\section{\bf Form factors and $D^{+}\rightarrow \sigma \pi ^{+}$ and 
$D^{+}\rightarrow \rho ^0\pi ^{+}$ decays}

The valence quark contribution shown in Fig. 1 gives 
\begin{equation}
J^{(\sigma)}_\mu=\int\frac{d^3 K}{\left( 2\pi \right) ^3}F_\mu^{(\sigma)} ,
\label{13} \end{equation}
where $F^{(\sigma)}_\mu $ is the matrix element 
\begin{eqnarray}
F^{(\sigma)}_\mu  &=&-ig_{\sigma q\bar{q}}\sqrt{\frac{m_d}{p_{d_0}}}\bar{v}
^i\left( p_d\right) \left( I\right) _i^j\frac{\left( p\hspace{-0.2cm}
/_d^{\prime }+m_d\right) _j^k}{p_d^{\prime 2}-m_d^2}\left( \gamma _\mu
\gamma _5\right) _k^lu_l\left( p_c\right) \sqrt{\frac{m_c}{p_{c_0}}} 
\nonumber \\
&&\times \left( \sqrt{2m_D}\frac 1{\sqrt{2}}\sqrt{3}\bar{u}^m\left(
p_c\right) \left( \gamma _5\right) _m^n v_n\left( p_d\right) \phi _D\left( 
{\bf K}\right) \right)\hskip 2pt . \label{14} 
\end{eqnarray}
Here, the term within the parenthesis is the bound state wave function of 
the $D$-meson, $\sqrt{3}$ being the color factor. We define the kinematical 
variables ${\bf K}={\bf p}_c-{\bf p}_d$ and ${\bf P}={\bf p}_c+{\bf p}_d$, 
so that ${\bf K}$ is the relative momentum and ${\bf P}$ is the center of 
mass momentum of the $c\bar{d}$ system. 

The evaluation of the trace implied in Eq.~(\ref{14}) gives: 
\begin{eqnarray}
F_\mu ^{(\sigma )} &=&-i4C({\bf K})g_{\sigma q\bar{q}}\{(p_c\cdot
p_d+m_cm_d)p_{d\mu }^{\prime }  \nonumber \\
&&-(p_d^{\prime }\cdot p_c+m_cm_d)p_{d\mu }+(p_d^{\prime }\cdot
p_d-m_d^2)p_{c\mu }\}\frac 1{p_d^{\prime 2}-m_d^2},
\label{15}\end{eqnarray}
where 
\begin{equation}
C\left( {\bf K}\right) =\sqrt{2m_D}\frac 1{\sqrt{2}}\sqrt{3}\sqrt{\frac{
m_dm_c}{p_{d_0}p_{c_0}}}\frac 1{4m_cm_d}\phi _D\left( {\bf K}\right)
\hskip 2pt .\label{16} \end{equation}
Working in the $D$-meson rest frame $\left( {\bf P}=0\right) $, where 
\begin{equation}
p_d^{\prime 2}-m_d^2=-\frac{m_D^2-m_c^2+m_d^2}2\left( 1-\frac{q^2}{m_D^2}
\right) +\frac{m_D^2+m_c^2-m_d^2}2\frac{k^2}{m_D^2}+{\bf q}\cdot {\bf K}
\hskip 2pt ,\label{17} \end{equation}
and noting that, if $\phi _D\left( {\bf K}\right) $ is of Gaussian type, $
{\bf K\simeq 0}$ dominates in the integration \cite{isgur}, one obtains \cite
{details}
\begin{eqnarray}
J_\mu ^{(\sigma )} &=&4iC\left( 0\right) g_{\sigma q\bar{q}}\frac{
m_D^2-\left( m_c-m_d\right) ^2}{2m_D^2}\frac 1{m_D^2-m_c^2+m_d^2}\left\{ 
\frac 1{1-\frac{q^2}{m_D^2}-a\frac{k^2}{m_D^2}}\right\}   \nonumber \\
&&\times \left\{ \left( m_D^2-2m_d\left( m_c+m_d\right) \right) \left(
p+k\right) _\mu -\left( m_D^2+2m_d\left( m_c+m_d\right) \right) q_\mu
\right\} ,
\label{18}\end{eqnarray}
where
\[
a=\frac{m_D^2+m_c^2-m_d^2}{m_D^2-m_c^2+m_d^2}\hskip 2pt .
\]
\par\noindent 
Note that, in the above approximation, 
${\displaystyle{4\pi\int K^2\hskip 2pt dK\hskip 2pt \phi_D (K )}}$ 
becomes ${\displaystyle{\int d^3 K\phi_D ({\bf K})}}$, which is the 
Fourier transform of the wave function at the origin, and we write it 
as $\phi_D(0)$ or equivalently $C(0)$. 

To eliminate $4C\left( 0\right)$, we consider the matrix element 
\begin{equation}
\left\langle 0\left| A_\mu \right| D\left( p\right) \right\rangle
=if_Dp_\lambda
\label{19} \end{equation}
which, when evaluated in the same valence quark approximation employed for 
the calculation of $J_\mu ^{(\sigma )}$, gives: 
\begin{equation}
f_D=\frac{4C(0)}{2m_D^2}(m_c+m_d)\left[ m_D^2-(m_c-m_d)^2\right]
\hskip 2pt .\label{20} \end{equation}

Thus, we finally obtain the valence quark triangle contribution 
\begin{equation}
G_{+}\left( q^2\right) =g_{\sigma q\bar{q}}\left( \frac{f_D}{m_c+m_d}
\right) \frac{m_D^2-2m_d\left( m_c+m_d\right) }{m_D^2-m_c^2+m_d^2}\frac 1{1-
\frac{q^2}{m_D^2}-a\frac{k^2}{m_D^2}},  \label{21}
\end{equation}
and, for $k^2=m_\sigma ^2$, this gives: 
\begin{equation}
G_{0}^{D\sigma }\left( m_\pi^2\right)\simeq G_{+}\left( 0\right)
=g_{\sigma q\bar{q}}\left( \frac{f_D}{m_c+m_d}\right) \frac{m_D^2-2m_d\left(
m_c+m_d\right) }{m_D^2-m_c^2+m_d^2}\frac 1{1-a\frac{m_\sigma ^2}{m_D^2}}.
\label{22}
\end{equation}

An exactly similar calculation for the case of the $\rho ^0$ in the 
$\rho$-dominance approximation ($k^2=0$), so that $g_{\rho d\bar{d}} 
\frac{f_\rho }{\sqrt{2}m_\rho ^2}=-\frac 12$, gives: 
\begin{equation}
-iq^\mu J_\mu ^{(\rho)} \equiv 
\left( q\cdot \epsilon ^{*}\right) \left( 2m_\rho
\right) A_0\left( q^2\right) =-\frac{\sqrt{2}m_\rho ^2}{2f_\rho }f_D\frac{1+
\frac{q^2}{m_D^2}}{1-\frac{q^2}{m_D^2}}q\cdot \epsilon ^{*}.  \label{23}
\end{equation}
Thus, for $q^2\simeq m_\pi ^2\simeq 0$, on using the KSRF relation $f_\rho =
\sqrt{2}f_\pi m_\rho $ \cite{ksrf}, we find: 
\begin{equation}
A_0^{D\rho }\left( 0\right) =-\frac 14\left( \frac{f_D}{f_\pi }\right) .
\label{24} \end{equation}
Note that this result is independent of quark masses in contrast to 
Eq.~(\ref{22}). It is, however, subject to a suppression factor $F_\rho
\left( 0\right) $ due to the off-mass-shellness of the $\rho$-meson 
$\left[
F_\rho \left( m_\rho ^2\right) =1\right] $. From the $\rho $-dominance of
the pion form factor, the experimental determination $\gamma _{\rho \pi \pi }
\frac{f_\rho }{m_\rho ^2}=1.22\pm 0.03$ indicates $F_\rho \left( 0\right)
\simeq 0.8$ \cite{riaz}. Accordingly, we rewrite Eq.~(\ref{24}) as: 
\begin{equation}
A_{0}^{D\rho }\left( 0\right) =-\frac 14\left( \frac{f_D}{f_\pi }
\right) F_\rho \left( 0\right) =-1.52f_D\text{ GeV}^{-1}.  \label{25}
\end{equation}

To account for the effect of the $a_2$-term in Eq.~(\ref{8}), we use the KSRF 
relation and the numerical value \cite{21}, 
\begin{equation}
F_{+}^{D\pi }\left( m_\rho ^2\right) \simeq \frac{F_{+}^{D\pi}
\left( 0\right)}{
\left(1-\frac{m^2_\rho}{m^2_D}\right)
\left(1-\frac{m^2_\rho}{m^2_{D'}}\right)}
\simeq \frac{1}{4}\left( 1.62 \right) \frac{f_D}{f_\pi },
\label{26}\end{equation}
with $F_{+}^{D\pi}(0) \simeq 0.3 \frac{f_D}{f_\pi }$, 
$\frac{m_{D'}}{m_D}=1.14$, $D'$ being the radial excitation of the $D$. 
Using 
$f_D= 0.23$ GeV \cite{burdman}, $F_{+}^{D\pi}(0)=0.53$, not inconsistent 
with its other estimates \cite{BABAR}. With $-a_2/a_1=0.44$, the 
square bracket on the right-hand side of Eq.~(8) has the value 
\begin{equation} 
\left[A_{0}^{D\rho}\left( 0\right)+0.44F_+^{D\pi}
\left( m^2_\rho\right) 
\right]\simeq -\left[ F_\rho \left( 0\right) -0.71\right] 
\frac{1}{4}\frac{f_D}{f_\pi }
=-0.09\times \frac{1}{4}\frac{f_D}{f_\pi}. 
\label{27}\end{equation}
This indicates that, in the framework used here, the $a_2$-term of Eq.~(1) 
can give a significant contribution to the $D\rightarrow \rho \pi $ channel. 

To obtain the numerical estimate for $G_0^{D\rho }\left( 0\right) $ from 
Eq.~({\ref{22}), we have to first fix $g_{\sigma q\bar{q}}$. 
The linear $\sigma $-model gives \cite{dib,ebert,scadron}:
\begin{eqnarray}
v &=&\left\langle \sigma \right\rangle =\frac{f_\pi }{\sqrt{2}};\qquad 
g = g_{\sigma qq}=g_{\pi qq}; \qquad 
g_{\sigma \pi \pi } =2\lambda v=2g^{\prime };  \nonumber \\
m_\sigma ^2 &=&2\lambda v^2; \qquad\quad\ \  
m_q = gv = g\frac{f_\pi }{\sqrt{2}}; \qquad\quad
g^{\prime } = 2gm_q=\sqrt{2}g^2f_\pi\hskip 2pt .   \label{28}
\end{eqnarray}
From these relations one finds: 
\begin{eqnarray}
g_{\sigma \pi \pi } &=&\frac{\sqrt{2}m_\rho ^2}{f_\pi }=2g^{\prime }
\label{29} \\
g_{\sigma qq} &=&g=\left( \frac{g^{\prime }}{\sqrt{2}f_\pi }\right)
^{1/2}=\left( \frac{g_{\sigma \pi \pi }}{2\sqrt{2}f_\pi }\right) ^{1/2}=
\frac{m_\sigma }{\sqrt{2}f_\pi }\simeq 2.57  \label{30} \\
m_q &=&240\text{ MeV}  \label{31}
\end{eqnarray}

Using Eqs. (\ref{30}) and (\ref{31}), $m_D=1.87 $ GeV and $m_c=1.45$ GeV, 
we obtain 
\begin{equation}
G_{0}^{D\sigma }\left( 0\right) =3.7f_D\text{ GeV}^{-1}  \label{32}
\end{equation}

The $a_1$-pole contribution from Fig. 2 gives, on using 
Eqs. (\ref{9}-\ref{11}): 
\begin{eqnarray}
\left\langle \sigma \left( k\right) \pi ^{+}\left( q\right) \left| H_{\text{
eff}}\right| D^{+}\left( p\right) \right\rangle  &=&\frac{G_F}{\sqrt{2}}
V_{cd}^{*}V_{ud}a_1\left( if_Df_{a_1}\right) p_\mu   \nonumber \\
&&\times \left[ -g^{\mu \lambda }+\frac{p^\mu p^\lambda }{m_{a_1}^2}\right] 
\frac{-1}{m_{a_1}^2-p^2}\frac i2\gamma _{a_1\sigma \pi }\left( q-k\right)
_\mu   \nonumber \\
&=&-\frac{G_F}{\sqrt{2}}V_{cd}^{*}V_{ud}a_1f_Df_{a_1}\gamma
_{a_1\sigma \pi }\frac{p\cdot \left( q-k\right) }{2m_{a_1}^2},  
\label{33} \\
\left\langle \rho \left( k\right) \pi ^{+}\left( q\right) \left| H_{\text{eff
}}\right| D^{+}\left( p\right) \right\rangle  &=&\frac{G_F}{\sqrt{2}}
V_{cd}^{*}V_{ud}a_1f_Df_{a_1}ip_\mu   \nonumber \\
&&\times \left[ -g^{\mu \lambda }+\frac{p^\mu p^\lambda }{m_{a_1}^2}\right] 
\frac{-1}{m_{a_1}^2-p^2}i\left( m_{a_1}^2-m_\rho ^2\right) f_{a_1\rho \pi
}\epsilon _\lambda ^{*}  \nonumber \\
&=&-\frac{G_F}{\sqrt{2}}V_{cd}^{*}V_{ud}a_1f_Df_{a_1}\left(
m_{a_1}^2-m_\rho ^2\right) f_{a_1\rho \pi }q\cdot \epsilon ^{*}.  \label{34}
\end{eqnarray}
Now $p\cdot \left( q-k\right) =m_\pi ^2-m_\sigma ^2$ independent of $p^2$, 
and the above equations give, in the square brackets on the right-hand sides 
of Eqs.~(\ref{7}) and (\ref{8}), the additional contributions to 
$G_{0}^{D\sigma}$ and $A_{0}^{D\rho}$, respectively:
\begin{eqnarray}
{\cal G}_{a_1}^{D\sigma } &=&-\frac{f_Df_{a_1}}{f_\pi }\frac{m_\pi
^2-m_\sigma ^2}{m_D^2-m_\sigma ^2}\frac 1{2m_{a_1}^2}\gamma _{a_1\sigma \pi }
\label{35} \\
{\cal A}_{a_1}^{D\rho } &=&-\frac{f_Df_{a_1}}{f_\pi }\frac{
m_{a_1}^2-m_\rho ^2}{2m_\rho m_{a_1}^2}f_{a_1\rho \pi }\hskip 2pt .  
\label{36} \end{eqnarray}

Moreover, the effective Lagrangian approach to Chiral symmetry gives \cite
{craigie}:
\begin{eqnarray*}
g_{a_1\rho \pi } &=&0,\hspace{1.0cm}f_{a_1\rho \pi }=\frac 1{\sqrt{2}f_\pi }
,\hspace{1.0cm}m_{a_1}=\sqrt{2}m_\rho , \\
f_{a_1} &=&f_\rho =\sqrt{2}f_\pi m_\rho ,\hspace{1.0cm}\gamma _{a_1\sigma
\pi }=\sqrt{2}\gamma _{\rho \pi \pi }=\sqrt{2}\frac{m_\rho }{f_\pi }.
\end{eqnarray*}

Using the above relations, we obtain for Eqs.~(\ref{35}) and (\ref{36}) 
the numerical values 
\begin{eqnarray}
{\cal G}_{a_1}^{D\sigma } &=&-\frac 12\frac{f_D}{f_\pi }\frac{m_\pi
^2-m_\sigma ^2}{m_D^2-m_\sigma ^2}=0.27f_D\text{ GeV}^{-1}  \label{37}
\\
{\cal A}_{a_1}^{D\rho } 
&=&-\frac 14\frac{f_D}{f_\pi }=-1.9f_D\text{GeV}^{-1}  \label{38}
\end{eqnarray}
and finally, using Eqs.~(\ref{27}), (\ref{32}), (\ref{37}) and (\ref{38}), 
the total contributions to the sqare brackets in the right-hand sides of 
Eqs.~(\ref{7}) and (\ref{8}) become: 
\begin{equation}
\left[G_0^{D\sigma }+{\cal G}_{a_1}^{D\sigma} \right] 
\simeq \left[ 1+0.073\right] 3.7f_D\text{ GeV}^{-1}
\hskip 2pt ,\label{39} 
\end{equation}
\begin{equation}
\left[ A_{0}^{D\rho } +0.44\hskip 2pt F_+^{D\pi}(m_\rho^2)+ 
{\cal A}_{a_1}^{D\rho} \right] 
\simeq -\left[ 0.09+1\right] \left( 1.9\right) 
f_D\text{GeV}^{-1}\hskip 2pt .  \label{40}
\end{equation}
For $f_D\simeq 230$ MeV, one gets 
\begin{equation}
\left[G_0^{D\sigma }+ {\cal G}_{a_1}^{D\rho}\right]    
\simeq 0.91\hskip 2pt ,  \label{41}\end{equation}
to be compared with $0.79\pm 0.15$ needed \citer{polosa,gatto} to explain 
the experimental branching ratio for 
$D^{+}\rightarrow \sigma \pi ^{+}$. Clearly, predicted branching 
ratios depend on the actual values of $f_D$ 
(and $f_B$) which, hopefully, will be experimentally determined in the near 
future \cite{thaler}. With the same values we obtain, from Eq.~(40), the 
width 
$\Gamma\left(D^{+}\rightarrow\rho^0\pi^{+}\right)=10.39\times 10^{-16}$ GeV
giving the branching ratio $1.66\times 10^{-3}$ to be compared with its 
experimental value $\left( 1.05\pm 0.31\right)\times 10^{-3}$ \cite{groom}. 

If we extend the previous analysis to $D_s\rightarrow \phi \pi $ where $\phi
\left( 1020\right) $ is treated as a pure $\bar{s}s$ state, we obtain $
\left[ \left\langle 0\left| \bar{s}\gamma _\mu s\right| \phi \right\rangle
=f_\phi \epsilon _\mu \right] $: 
\begin{equation}
A_0^{D\phi }\simeq \frac{f_{D_s}}{2m_\phi }\frac{m_\phi ^2}{f_\phi }.
\label{42}\end{equation} 
In this case the intermediate $a_1$-exchange should be absent and, in the 
factorization approximation, the $a_2$-term in $H_{\text{eff}}$ should not 
contribute. Using $f_\phi \simeq 0.23$ GeV$^2$ from $\Gamma \left( \phi
\rightarrow e^{+}e^{-}\right) $, we would obtain 
\begin{equation}
A_0^{D\phi }\simeq 2.2f_{D_s}\text{ GeV}^{-1}\simeq 0.62.
\label{43} \end{equation}
This leads to 
$\Gamma\left(D_s\rightarrow \phi \pi \right) \simeq
2.8\times 10^{-14}$ GeV and ${\cal B}\left( D_s\rightarrow \phi \pi \right)
\simeq 2.1\%$, compatible  
with the experimentally measured value $3.6\pm 0.9\%$ 
\cite{groom} and the theoretical estimate of 
Ref.~\cite{deandrea4}.\footnote{Treating the $f_0(980)$ as a pure 
$\bar{s}s$ state we would obtain from the 
analogous quark triangle diagram, with $m_s\approx 1.6 m_q$, a value 
for ${\cal B}(D_s\to f_0\pi )$ substantially larger than the experimental 
one (and the result of \cite{deandrea4}). To have agreement we would require 
a mixing angle with the nonstrange scalar-isoscalar component of the order 
of 10 - 20 degrees for $m_q=(0.24 - 0.31)$ GeV.
Thus, our model does not favour the description of $f_0$ as a pure $\bar{s}s$ 
state.}

\section{\bf $B\rightarrow \sigma \pi$, $B\rightarrow \rho \pi $ decays}

The effective weak Hamiltonian is given by \cite{bauer} 
\begin{equation}
H_{\text{eff}}=\frac{G_F}{\sqrt{2}}V_{ub}^{*}V_{ud}\left\{ a_1\left( \bar{u}
b\right) _{\text{V}-\text{A}}\left( \bar{d}u\right) _{\text{V}-\text{A}
}+a_2\left( \bar{d}b\right) _{\text{V}-\text{A}}\left( \bar{u}u\right) _{
\text{V}-\text{A}}\right\} ,  \label{44}
\end{equation}
where the Wilson coefficients $c_1$ and $c_2$, fitted for $B$-decays, are $
c_1\left( m_b\right) =1.105$ and $c_2\left( m_b\right) =-0.228$ so that $
a_1=c_1+\frac 13c_2=1.03$ and $a_2=c_2+\frac 13c_1=0.14.$ The factorization
ansataz gives for the decay $B^{-}\to \sigma \pi ^{-}$ the analogue of 
Eqs.~(\ref{21}) and (\ref{22}). With $m_B=5.28$ GeV, $m_b=4.757$ GeV, 
$m_d=0.240$ GeV, one obtains: 
\begin{equation}
\left\langle \sigma\left( k\right) \pi ^{-}\left( q\right) \left| H_{
\text{eff}}\right| B^{-}\left( p\right) \right\rangle =\frac{G_F}{\sqrt{2}}
V_{ub}^{*}V_{ud}a_1f_\pi \left( m_B^2-m_\sigma ^2\right) 
\left[ G_0^{B\sigma}\left( m_\pi ^2\right) \right],  \label{45}
\end{equation}
and the valence quark triangle contribution
\begin{equation}
G_{0}^{B\sigma }=2.67f_B\text{ GeV}^{-1}.  \label{46}
\end{equation}
With $f_B=0.150$ GeV, this gives [the $a_1$-pole contribution is negligible 
because of the factor $\left( m_\sigma ^2/m_B^2\right) /\left( 1-m_\sigma
^2/m_B^2\right) $ in Eq.~(37)]: 
\begin{equation}
G_0^{B\sigma }=0.4,
\label{47} \end{equation}
consistent with the value found in \cite{deandrea}.

For $B\rightarrow \rho \pi $ decays, using the factorization ansatz: 
\begin{eqnarray}
&&\left\langle \rho ^0\left( k\right) \pi ^{-}\left( q\right) \left| H_{
\text{eff}}\right| B^{-}\left( p\right) \right\rangle  \nonumber
 \\
&=&\frac{G_F}{\sqrt{2}}V_{ub}^{*}V_{ud}\left[ a_1f_\pi \left( -iq^\mu
\right) \left\langle \rho ^0\left( k\right) \left| A_\mu \right| B^{-}\left(
p\right) \right\rangle +a_2\left( \frac{f_\rho }{\sqrt{2}}\right) \epsilon
^{*\mu }\left\langle \pi ^{-}\left( q\right) \left| V_\mu \right|
B^{-}\left( p\right) \right\rangle \right]   \nonumber \\
&=&\frac{G_F}{\sqrt{2}}V_{ub}^{*}V_{ud}\left[ a_1f_\pi \left( 2m_\rho
\right) \epsilon ^{*}\cdot qA_0^{B\rho ^0}\left( m_\pi ^2\right) +a_2\left( 
\frac{f_\rho }{\sqrt{2}}\right) \left( 2\epsilon ^{*}\cdot {q}\right)
F_{+}^{B^{-}\pi ^{-}}\left( m_\rho ^2\right) \right],   \label{48}
\end{eqnarray}
\begin{equation}
\left\langle \rho ^{+}\left( k\right) \pi ^{-}\left( q\right) \left| H_{
\text{eff}}\right| \bar{B}^0\left( p\right) \right\rangle 
=\frac{G_F}{\sqrt{2}}V_{ub}^{*}V_{ud}a_1f_\pi \left( 2m_\rho \right)
\epsilon ^{*}\cdot q \left[ A_0^{\bar{B}^0\rho ^{+}}\left( m_\pi ^2\right)
\right] , \label{49}
\end{equation}
\begin{eqnarray}
\left\langle \rho ^{-}\left( k\right) \pi ^{+}\left( q\right) \left| H_{
\text{eff}}\right| \bar{B}^0\left( p\right) \right\rangle 
&=&\frac{G_F}{\sqrt{2}}V_{ub}^{*}V_{ud}a_1f_\rho \epsilon ^{*\mu
}\left\langle \pi ^{+}\left( k\right) \left| A_\mu \right| \bar{B}^0\left(
p\right) \right\rangle  \nonumber \\
&=&\frac{G_F}{\sqrt{2}}V_{ub}^{*}V_{ud}a_1\left( \sqrt{2}f_\pi m_\rho
\right) \left( 2\epsilon ^{*}\cdot q\right) 
\left[F_{+}^{\bar{B}^0\pi ^{+}}\left(m_\pi ^2\right)\right] .\label{50}
\end{eqnarray}
Here: $A_\mu ={\bar{u}}\gamma _\mu \gamma _5b$, $V_\mu ={\bar{u}}\gamma _\mu
b$, and 
\begin{equation}
\left\langle \pi ^{+}\left( k\right) \left| V_\mu \right| \bar{B}^0\left(
p\right) \right\rangle =\left( p+q\right) _\mu F_{+}\left( k^2\right)
+\left( p-q\right) _\mu F_{-}\left( k^2\right) .  \label{51}
\end{equation}
Noting the relations 
\[
g_{\rho ^{+}\bar{u}d}=\sqrt{2}g_{\rho ^0u\bar{u}}=\frac{m_\rho ^2}{f_\rho }=
\frac{m_\rho }{\sqrt{2}f_\pi },
\]
the quark triangle diagrams give: 
\begin{equation}
A_{0}^{\bar{B}^0\rho ^{+}}=\sqrt{2}A_{0}^{B^{-}\rho ^0}=
\frac{\sqrt{2}}4\frac{f_B}{f_\pi }=\sqrt{2}\left( 0.25\right) \frac{f_B}{
f_\pi }\hskip 2pt .  \label{52}
\end{equation}

The form factor $F_{+}^{\bar{B}^0\pi }$ introduced in Eq.~(\ref{51}) 
has been found to be about 0.30 \cite {BABAR,details}, so that, with 
$f_B=0.150$ GeV [notice that, here, 
$F_+^{B\pi} (m^2_\rho) \simeq F_+^{B\pi} (0)$ to a very good approximation 
as $m^2_\rho / m^2_B$ corrections are negligible]:
\begin{equation}
F_{+}^{\bar{B}^0\pi ^{+}}\left( 0\right) =F_{+}^{B^{-}\pi ^{-}}\left(
0\right) \simeq 0.26\frac{f_B}{f_\pi }.  \label{53}
\end{equation}

Now, the $a_1^{-}$-pole contributes to $B^{-}\rightarrow \rho ^0\pi ^{-}$
and, in vacuum saturation, negligibly to 
$\bar{B}^0\rightarrow \rho ^{\pm }\pi ^{\mp }$, the latter 
contribution being controlled by the small $a_2$ coefficient. This can 
enhance the branching ratio for $B^{-}\rightarrow \rho ^0\pi ^{-}$ and, 
as such, provide a mechanism (in addition to the $\sigma $-contribution to $
B^{-}\rightarrow \rho ^0\pi ^{-}$ decay \cite{deandrea}) to lower the ratio 
${\cal R}$. The additional, intermediate $a_1$-contribution to be included 
in the square brackets on the right-hand sides of Eqs.~(\ref{48})-(\ref{50}), 
see Eq.~(\ref{38}), is given by 
\begin{equation}
{\cal A}_{a_1}^{B^{-}\rho ^0}=\left( 0.25\right) \frac{f_B}{f_\pi }.
\label{54}\end{equation}
One can note the change of sign since the 
$a_1^{-}\rightarrow \rho ^0\pi ^{-}$ coupling has sign opposite to 
$a_1^{+}\rightarrow \rho ^0\pi ^{+}$, and similar is the case for 
the relative signs of $a_1^0\rightarrow \rho ^{+}\pi ^{-}$ and 
$a_1^0\rightarrow \rho^{-}\pi ^{+}$. Thus, on using 
Eqs.~(\ref{48})-(\ref{54}), and the suppression factor 
$F_\rho(0)\simeq 0.8$ to take care of the off-mass-shellness of the 
$\rho$-meson in Eq.~(\ref{52}), one finds [we also include the small 
contribution controlled by $a_2/a_1\simeq 0.13$ of the $a_1$ meson 
to the $\rho^{\pm}\pi^{\mp}$ modes]:
\begin{equation}
{\cal R}=\left( \sqrt{2}\right) ^2\frac{\left[ 0.20+0.25\cdot 0.13/
\sqrt{2}\right]^2+\left[ 0.26-0.25\cdot 0.13/\sqrt{2}\right]^2}{
\left[ 0.20+0.26\cdot 0.13+0.25\right]^2}\approx 0.91\hskip 2pt , 
\label{55}
\end{equation}
in the lower range, but still consistent with the interval allowed by the 
experimental determination. Note that this ratio is almost 
independent of the value of $f_B/f_\pi $, and that the effect of the 
$a_2$-term of Eq.~(\ref{44}) is 
almost negligible.\footnote{Actually, in principle the ``long-distance'' 
$a_1$-meson contribution could be subject to a suppression factor taking 
into account the $a_1$ off-mass-shellness. This effect does not relate to 
the $a_1$-meson propagator, that is cancelled by a corresponding numerator, 
see Eqs.~(\ref{33})-(\ref{36}), but may reside in the coupling 
$f_{a_1\rho \pi }$. Such correction might be taken into account by 
introducing a $B$-factor $B_{a_1}$. Assuming $B_{a_1}\simeq 0.7-0.8$, 
{\it i.e.}, the same order of magnitude found for  
$K-{\bar K}$ and $B-{\bar B}$ mixing \cite{BABAR}, the correction 
would slightly increase the numerical result for ${\cal R}$ in 
Eq.~(\ref{55}), thus improving the agreement with the experimental value.}
The individual branching ratio is 
\[
{\cal B}\left( B^{-}\rightarrow \rho ^0\pi ^{-}\right) =1.99\left|
V_{ub}\right| ^2=\left( 2.43\pm 2.08\right) \times 10^{-5}
\]
for $\left| V_{ub}\right| =\left( 3.5\pm 1.5\right) \times 10^{-3}$, 
that is compatible, within the uncertainty, with the experimental upper limit 
${{\cal B}<(1.0\pm 0.4)\times 10^{-5}}$ \cite{groom}.

\section{Conclusions}

Our analysis of the decays $D^{+}\rightarrow \sigma \pi ^{+},$ $
D^{+}\rightarrow \rho ^0\pi ^{+},$ $B^{-}\rightarrow \sigma \pi ^{-},$ $
B^{-}\rightarrow \rho ^0\pi ^{-}$ and $\bar{B}^0\rightarrow \rho ^{\pm }\pi
^{\mp }$show that the valence quark ``triangle'' graph, supplemented by the 
long distance $a_1$-exchange, is in reasonable agreement with the 
available branching ratios, in particular with 
$D^{+}\rightarrow \rho ^0\pi ^{+}$ and that of 
$D^{+}\rightarrow \sigma \pi ^{+}$ recently measured. The 
contribution from the $a_1$-pole has also been found important. In 
particular, the inclusion of this contribution gives the ratio 
\[
{\cal R}=\frac{{\cal B}\left( \bar{B}^0\rightarrow \rho ^{\pm }\pi ^{\mp
}\right) }{{\cal B}\left( \bar{B}^{-}\rightarrow \rho ^0\pi ^{-}\right) }
\approx 0.9, \]
consistent with the experimental values within the large experimental
uncertainities. More accurate determinations of this ratio would provide a 
stringent test of the model presented here. 

\bigskip
\bigskip
\leftline{\bf Acknowledgement}
\par\noindent
One of the authors (R) would like to thank Professor M. A. Virasoro for 
hospitality at the Abdus Salam International Center for Theoretical Physics, 
Trieste where most of this work was done. NP acknowledges partial financial 
support from MURST (Italian Ministry of University, Scientific Research and 
Technology) and from funds of the University of Trieste.

\vfill



\newpage

\begin{figure}[tbp]
\centerline{\epsfig{figure=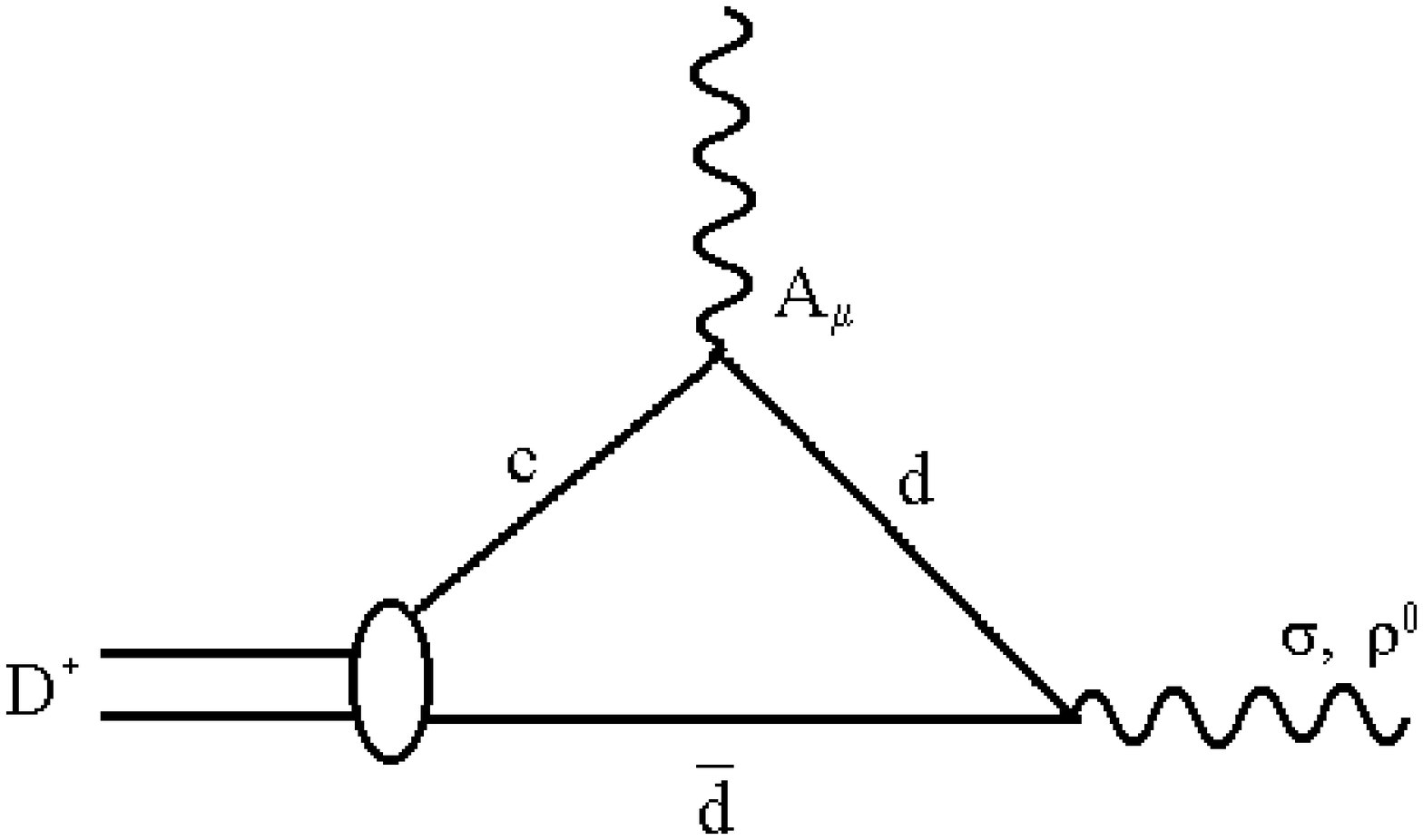,height=7cm}}
\caption{Quark triangle graph for $\left\langle \sigma ,\rho ^0\left|
d\gamma _\mu \gamma _5c\right| D^{+}\right\rangle $}
\end{figure}
\bigskip
\begin{figure}[tbp]
\centerline{\epsfig{figure=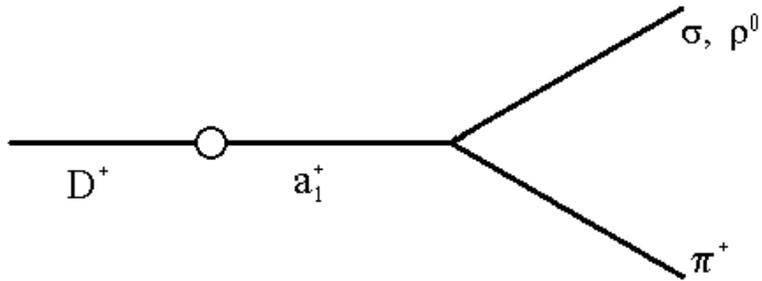,height=5cm}}
\caption{$a_1$-pole contribution to $D^{+}\rightarrow \sigma \left( \rho
^0\right) \pi ^{+}$}
\end{figure}

\end{document}